\documentclass[conference]{IEEEtran}
\usepackage{blindtext, graphicx,booktabs}
\usepackage[top=1in,bottom=0.75in,right=0.75in,left=0.75in]{geometry}
\usepackage{amsmath,amssymb}
\usepackage{subfigure}

\title{\LARGE \bf
The Impact of Road Configuration on V2V-based Cooperative Localization
}

\begin{document}

\author{\IEEEauthorblockN{ }
\IEEEauthorblockA{Macheng Shen \\Department of Naval Architecture \\
 and Marine Engineering\\
University of Michigan \\
Ann Arbor, Michigan 48109\\
Email: macshen@umich.edu }
\and
\IEEEauthorblockN{ }
\IEEEauthorblockA{Ding Zhao \\
 Transportation Research Institute\\
 University of Michigan\\
 Ann Arbor, Michigan 48109\\
 Email: zhaoding@umich.edu
 }
\and
\IEEEauthorblockN{ \\  }
\IEEEauthorblockA{ Jing Sun\\
Department of Naval Architecture \\
 and Marine Engineering\\
University of Michigan \\
Ann Arbor, Michigan 48109\\
Email: jingsun@umich.edu
 }}

\maketitle

\begin{abstract}

Cooperative localization with map matching has been shown to reduce Global Navigation Satellite System (GNSS) localization error from several meters to sub-meter level by fusing the GNSS measurements of four vehicles in our previous work. While further error reduction is expected to be achievable by increasing the number of vehicles, the quantitative relationship between the estimation error and the number of connected vehicles has neither been systematically investigated nor analytically proved. In this work, a theoretical study is presented that analytically proves the correlation between the localization error and the number of connected vehicles in two cases of practical interest. More specifically, it is shown that, under the assumption of small non-common error, the expected square error of the GNSS common error correction is inversely proportional to the number of vehicles, if the road directions obey a uniform distribution, or inversely proportional to logarithm of the number of vehicles, if the road directions obey a Bernoulli distribution. Numerical simulations are conducted to justify these analytic results. Moreover, the simulation results show that the aforementioned error decrement rates hold even when the assumption of small non-common error is violated.

\end{abstract}

\section{INTRODUCTION}

Low cost Global Navigation Satellite Systems (GNSS) are used for most mobile applications, whose localization accuracy are typically in the range of several meters. Improving the localization accuracy of these widespread GNSS without incurring additional hardware and infrastructure costs has motivated recent research activities on Cooperative Map Matching (CMM). CMM has been shown able to improve Global Navigation Satellite System (GNSS) positioning of a group of connected vehicles through estimation and correction of the common GNSS localization error. Since the error caused by atmospheric delay and satellite clock error is almost the same to all the vehicles in the same area, this common error can be estimated by matching the vehicle positioning results to a digital road map, assuming that all the vehicles travel on lanes. With a properly estimated common error, the positioning of vehicles can be corrected, thus improving the localization accuracy. The improvement would largely depend on the quality of the common error estimation, which is determined by the CMM algorithm and the configuration of the road constraints.\\
\indent Recently, two different CMM algorithms have been developed for GNSS common error estimation problem, i.e., a non-Bayesian particle-based approach in Rohani et. al. \cite{r1} and a Bayesian approach based on a Rao-Blackwellized Particle Filter in our previous work \cite{r2}, \cite{r3}. \\
\indent In contrast, the effects of road configuration on the common error estimation have not been reported in open literature. Intuitively, in order to produce a good estimation of the common error, the road constraints should be rich enough so that common error in different directions can be detected. This richness is expected to be enhanced with the number of vehicles and the diversity of the road directions.\\
\indent In this work, the correlation between the estimation quality of the common error and the richness of the road constraints is quantified analytically. More specifically, the functional relationships between the mean square error of the common error estimation and the number of connected vehicles are derived under two different assumptions about road configurations. The results provide a guideline for the implementation of CMM and a foundation for the development of algorithms that intelligently select vehicles to include for maximal error reduction.\\
\indent In the following sections, details of the derivation of the error bounds are presented with justification through Monte Carlo simulations. In Section 2, an analytic expression of the estimation error as a function of the road configuration and the non-common error is derived. In Section 3, asymptotic formulas of the expectation of the estimation error with respect to Gaussian distributed non-common error are derived for uniformly or Bernoulli distributed road directions. Nonetheless, most of the results are derived under the assumptions of large number of vehicles and small non-common error. In Section 4, simulation results are presented to demonstrate and justify the applicability of the theoretical results to realistic scenarios where both the number of connected vehicles and the non-common error are finite. In Section 5, the contributions and conclusions are summarized.
\section{Estimation error of the common error}
In this section, we propose a framework of vehicle positioning within a reference road framework to facilitate the analytic investigation. The GNSS measurement will be decomposed into four components and one estimator of the common error will be formulated and the corresponding estimation error will be represented as an integral depending on the non-common error and the road constraints.\\
\begin{figure}[tb]
  \centering
  \includegraphics[width=0.49\textwidth]{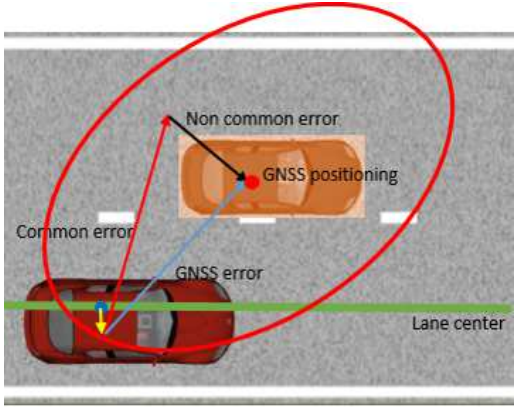}   
  \caption{Illustration of the deviation from the lane center (yellow arrow), the common error (red arrow), the non-common error (black arrow), the lane center point (blue dot) and the GNSS positioning (red dot)}
  \label{notation}
\end{figure}
\indent The coordinate of GNSS positioning of the $i$-th vehicle can be decomposed into superposition of the coordinate of a point on the corresponding lane center, the deviation of the vehicle from the lane center (yellow arrows), the common error (red arrows) and the non-common error (black arrows) as illustrated in Fig. \ref{notation}. The blue and red dots represent the true positions and the GNSS positioning, respectively. This relationship can be expressed mathematically as
\begin{equation}
x^G_i=x^L_i+x^D_i+x^C+x^N_i, i=1,2,...,N,
\end{equation}
where $x^G_i$ is the GNSS positioning of the $i$-th vehicle, $x^L_i$ is the closet point on the center of the lane from the vehicle, $x^D_i$ is the deviation of the vehicle coordinates from $x^L_i$, $x^C$ is the GNSS common localization error, $x^N_i$ is the GNSS non-common localization error including receiver noise error and multipath error and $N$ is the number of connected vehicles.\\
\indent The fact that all the vehicles travel on the roads can be expressed as a set of inequalities
\begin{equation}
g_i(x^G_i-x^C-x^N_i)<0.
\end{equation}
\indent In reality, the geometry of the road sides can be so complicated that the constraints cannot be expressed analytically. But frequently, the road sides can be approximated as straight lines such as those shown in Fig. \ref{notation}. In these cases, the constraint functions $g_i$ have simple analytic forms
\begin{equation}
g_i(x)=(x-x^L_i)\cdot n_i-w,
\end{equation}
where $\{\cdot\}$ is the dot product operator, $n_i$ is the unit vector normal to the lane center point towards outside of the road and $w$ is the half width of the lane. \\
\indent Alternatively, (2) can be interpreted as the feasible set of the common error given the GNSS positioning and the non-common error. The non-common error is unknown, however, to the implementation of CMM. Thus, an approximation of the feasible set by neglecting the non-common error is used instead of the exact feasible set, which is
\begin{equation}
\begin{aligned}
\Omega &=\{\tau|\bigcap\limits_{i=1}^{N}g_i(x^G_i-\tau)\}\\
&=\{\tau|\bigcap\limits_{i=1}^{N}g_i(x^L_i+x^C+\tilde{x}^N_i-\tau)\}\\
&=\{\tau|\bigcap\limits_{i=1}^{N}\tilde{g}_i(x^C+\tilde{x}^N_i-\tau)\},
\end{aligned}
\end{equation}
where 
\begin{equation}
\tilde{x}^N_i \triangleq x^D_i+x^N_i
\end{equation}
and 
\begin{equation}
\tilde{g}_i(x)\triangleq g_i(x+x^L_i)=x\cdot n_i-w
\end{equation}
A point estimator of the common error can be taken as the average over the approximate feasible set $\Omega$,
\begin{equation}
\hat{x}^C=\frac{1}{S}\int\limits_\Omega \tau dA, S=\int\limits_\Omega dA,
\end{equation}
where $\tau$ is the dummy integration variable and $dA$ is the area element.\\
\indent The estimation error of the common error is of practical interest, which can be evaluated as
\begin{equation}
\begin{aligned}
e&=x^C-\hat{x}^C\\
&=x^C-\frac{1}{S}\int\limits_\Omega \tau dA\\
&=\frac{1}{S}\int\limits_\Omega (x^C-\tau) dA\\
&=\frac{1}{S}\int\limits_{\Omega'} \tau' dA,
\end{aligned}
\end{equation}
where 
\begin{equation}
\tau'=x^C-\tau,
\end{equation}
and
\begin{equation}
\Omega'=\{\tau'|\bigcap\limits_{i=1}^{N}\tilde{g}_i(\tilde{x}^N_i+\tau')<0\}.
\label{eq8}
\end{equation}
\indent Eq. (8) and (10) states that the estimation error equals to the geometric center of the intersection of the road constraints perturbed by the composite non-common error $\tilde{x}^N_i$. \\
\indent It has been shown by experimental data that the GNSS non-common error can be well approximated as Gaussian random variable in \cite{r2}. The Safety Pilot dataset collected in Ann Arbor shows that statistically the deviations from the lane center also obey Gaussian distribution. As a result, the composite error $\tilde{x}^N_i$ is a Gaussian random variable.\\
\indent The expectation of the square estimation error is of practical interest. In two special cases, analytic approximations to this expectation valid for large number of vehicles can be established.

\begin{figure*} \centering 
\subfigure[Orthogonal road model] { \label{fig:a} 
\includegraphics[width=0.8\columnwidth]{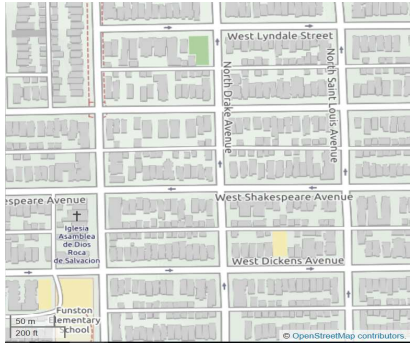} 
} 
\subfigure[Random road model] { \label{fig:b} 
\includegraphics[width=0.86\columnwidth]{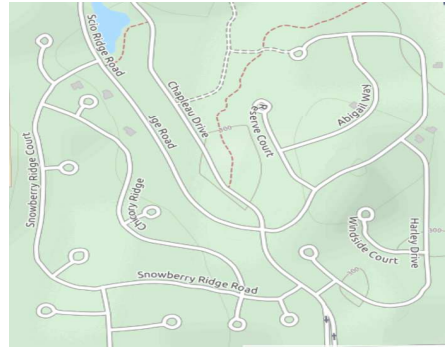} 
} 
\caption{Examples of road configurations that can be modeled by the presented two road angle distributions, images from OpenStreetMap} 
\label{fig} 
\end{figure*}

\section{Asymptotic Analysis for the expected square error}
In this section, we derive the asymptotic decay of the mean square error with respect to the number of vehicles in two typical cases of the road configuration, whose examples are shown in Fig. \ref{fig}. In the first case, the road configuration is modeled as cross roads where the roads are either parallel or orthogonal. In the second case, it is modeled as randomly oriented roads where the direction angles obey a uniform distribution.
\subsection{Orthogonal road directions}
In the first case, it is assumed that each road is parallel to one of the two orthogonal axes of the global reference frame. As a result, the direction angles $\theta$ of the vehicles relative to the reference frame belong to a set with four elements:
\begin{equation}
\theta_i\in \{0,\frac{\pi}{2}, \pi,\frac{3\pi}{2}\}
\end{equation}
\indent This case can be viewed as a simplified model for the urban areas where most roads are orthogonal to each other.\\
\indent Invoking (8), the square error can be expressed analytically as
\begin{equation}
e^2=\frac{X^2_1+X^2_2+X^2_3+X^2_4-2X_1X_3-2X_2X_4}{4},
\end{equation}
where $X_j,j=1,2,3,4$ are the largest projections of the composite non-common error on each of the four normal vector:
\begin{equation}
X_j=max\{\tilde{x}^N_{j_1}\cdot n_{j_1},\tilde{x}^N_{j_2}\cdot n_{j_2},...,\tilde{x}^N_{j_{Nj}}\cdot n_{j_{Nj}}\}, j=1,2,3,4.
\end{equation}
$N_j, j=1,2,3,4$ are the numbers of vehicles traveling in each of the four directions.\\ 
\indent If all the $\tilde{x}^N_i\cdot n_i, i=1,2,...N$ are independent and identically distributed, then according to the Fisher-Tippett-Gnedenko theorem \cite{r4}, the limit distribution of $X_j$ for large $N_j$ is Gumbel distribution whose cumulative distribution function is given by
\begin{equation}
F(X_j)=exp(-exp(-(X_j-\mu_j)/\beta_j)).
\end{equation}
\indent Moreover, The leading order of the normalization constants $\mu_j$ and $\beta_j$ are related to the variance of the Gaussian distribution $\sigma$ through \cite{r5}
\begin{equation}
\mu_j\sim\sigma\sqrt{2log(N_j)},\beta_j\sim\sigma\frac{1}{\sqrt{2log(N_j)}}.
\end{equation}
\indent Using the property of Gumbel distribution, the expectation of $e^2$ with respect to $X_j$ can be evaluated:
\begin{equation}
\begin{aligned}
E_X[e^2]&=\frac{\pi^2}{24}\sum_{j=1}^4\beta^2_j+\frac{1}{4}[\mu_1-\mu_3+\gamma(\beta_1-\beta_3)]^2\\ 
&+\frac{1}{4}[\mu_2-\mu_4+\gamma(\beta_2-\beta_4)]^2,
\end{aligned}
\end{equation}
where $\gamma\approx 0.5772$ is the Euler-Mascheroni constant.\\
\indent With the assumption that all the four $N_j$ are large number of the same order and using the asymptotic formulas (15), it can be readily shown that the first term in (16) is of $O(\frac{1}{log(N_j)})$ and the following two terms are of $O(\frac{1}{N_jlog(N_j)})$. Thus, the leading order asymptotic approximation is
\begin{equation}
E_X[e^2]\sim\frac{\pi^2\sigma^2}{48}\sum_{j=1}^4\frac{1}{log(N_j)}.
\end{equation}

\subsection{Uniformly distributed random road directions}
\indent In the second case, each direction angle is assumed to be randomly distributed within $[0,2\pi)$.
In addition, it is assumed that the non-common error is small enough such that (8) can be linearized with respect to the non-common error:
\begin{equation}
e=e_0+\Delta e=e_0+\frac{C\tilde{X}}{S_0},
\end{equation}
where 
\begin{equation}
e_0=\frac{1}{S_0}\int\limits_{\Omega_0} \tau' dA,
\end{equation}
\begin{equation}
\Omega_0=\{\tau'|\bigcap\limits_{i=1}^{N}\tilde{g}_i(\tau')<0\},
\end{equation}
\begin{equation}
\tilde{X}=[\tilde{x}^N_1\cdot n_1,\tilde{x}^N_2\cdot n_2,...,\tilde{x}^N_N\cdot n_N]^T,
\end{equation}
and 
\begin{equation}
C=S_0\frac{\partial e}{\partial \tilde{X}}.
\end{equation}
$C$ is a $2\times N$ matrix whose components are related to the geometric quantities of the road constraints.\\
\indent The condition under which the linearization (18) is valid is
\begin{equation}
||\tilde{X}||_\infty\ll\frac{2\pi w}{N},
\end{equation}
where $w$ is the half width of the lane.\\
\indent With the assumption that each non-common error obeys independent Gaussian distribution with zero mean, i.e., $\tilde{X}\sim N(0_{N\times 1}, diag(\sigma^2_1,\sigma^2_2,...,\sigma^2_N))$, the expectation of the square error is
\begin{equation}
E_X[e^2]=e^2_0+\frac{1}{S^2_0}tr(L^TC^TCL),
\label{24}
\end{equation}
where $L=diag(\sigma_1,\sigma_2,...,\sigma_N)$ is the Cholesky decomposition of the joint Gaussian covariance matrix.\\
\indent Eq. (\ref{24}) is quite useful in practice as it implicitly provides a measurement of the road configurations. Given any road configurations specified by $e_0$ and $C$, the mean square error can be evaluated. An optimization technique can be developed to select the best road configuration so as to optimize the localization accuracy. On the other hand, the theoretical value of this equation is demonstrated through the derivation of asymptotic error decrement shown as follows. \\
\indent As both $e_0$, $S_0$ and $C$ depends on the road direction angles $\theta_i$, $E_X(e^2)$ is also a random variable.
It can be shown that the expectation of $E_X(e^2)$ with respect to $\theta_i, i=1,2,...,N$ is of $O(\frac{1}{N})$ (See Appendix). More specifically,
\begin{equation}
E_{\theta}[e^2_0]=\frac{2w^2}{9N}+o(\frac{1}{N})
\end{equation}
and
\begin{equation}
E_{\theta}[\frac{1}{S^2_0}tr(L^TC^TCL)]=\frac{3\sum_{i=1}^N \sigma^2_i}{2N^2}+o(\frac{1}{N})
\end{equation}

\section{Simulation Justification}
\indent In this section, simulation results are presented to justify the validity of the asymptotic formulas derived in Section 3. The expectations are calculated by averaging over 5000 samples of $e^2$. Each sample value for the orthogonal road case is calculated through two approaches. One approach is the analytic formula (12), and the other approach is a Monte Carlo integration where the proposal distribution is a two-dimensional uniform distribution. Besides, the number of vehicles in each direction is the same. For the uniformly random road angle case, each sample value of $e^2$ is calculated by the Monte Carlo integration. The number of samples to implement each Monte Carlo integration is 10000. The road half width is $w=2\mbox{ }m$ in the simulation.\\
\subsection{Small non-common error: $\sigma=0.3\mbox{ }m$}
\begin{figure}[tb]
  \centering
  \includegraphics[width=3.5in]{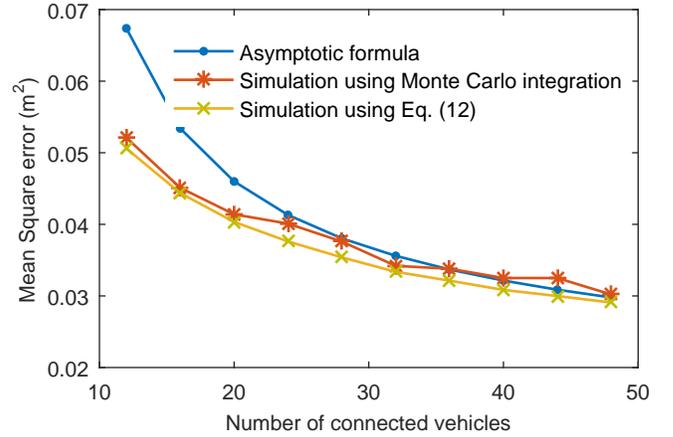}   
  \caption{Comparison between the asymptotic formula and numerical simulation results on the orthogonal road angle case}
  \label{f2}
\end{figure}

\begin{figure}[tb]
  \centering
  \includegraphics[width=3.5in]{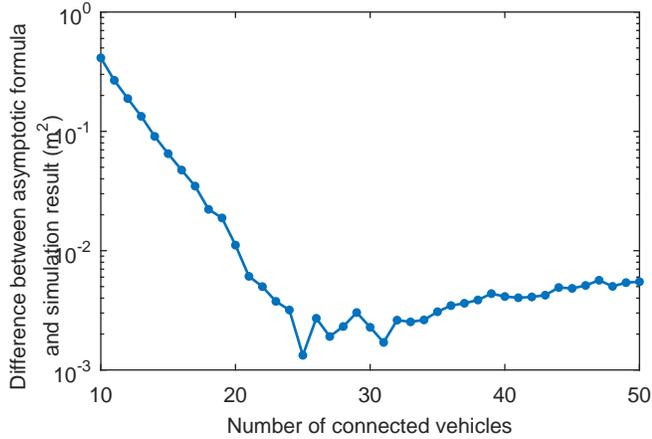}   
  \caption{Difference between the asymptotic formula and the numerical simulation result in Figure 3}
  \label{f4}
\end{figure}

\indent Fig. \ref{f2} shows the comparison of the orthogonal road case. The asymptotic formula is in good agreement with the numerical results as the number of vehicles increases. The results using the two numerical approaches are also different, which should be caused by the random error resulted from the Monte Carlo integration used to calculate $e^2$. Therefore, the one that uses (8) is expected to be closer to the underlying true expectation. Compared with this result, the asymptotic formula slightly overestimates the error. This difference may be the result of the fact that the convergence to the Gumbel distribution is rather slow \cite{r5}.\\
\begin{figure}[tb]
  \centering
  \includegraphics[width=3.5in]{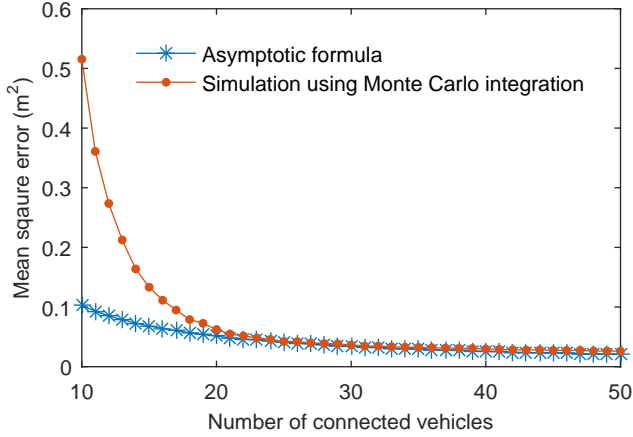}   
  \caption{Comparison between the asymptotic formula and numerical simulation results on the uniformly distributed random road angle case}
  \label{f3}
\end{figure}
\indent Fig. \ref{f3} shows the comparison of the random road angle case. Fig. \ref{f4} shows the corresponding difference between the asymptotic formula and the simulation results using Monte Carlo integration. The difference reaches its minimum around $N=25\sim 30$ and increases with the further increase of $N$. This result can be expected for the following two reasons. First, the asymptotic formula is derived for large $N$. As a result, the difference at small $N$ should be significant. Second, the linearization (18) based on the small non-common error assumption (23) eventually becomes invalid for fixed $\sigma$ and increasing $N$.
\subsection{Large non-common error: $\sigma=1\mbox{ }m$}
\indent The analytic results shown in Section 3 do not apply to the large non-common error case because the approximate feasible set described by (4) may be an empty set. Nonetheless, this problem can be addressed by assigning a weight to each hypothesis of the common error according to its compatibility with the road constraints. The weighted road map approach proposed by Rohani et. al. is applied to generate simulation results because of its simplicity for implementation.\\
\begin{figure}[tb]
  \centering
  \includegraphics[width=3.5in]{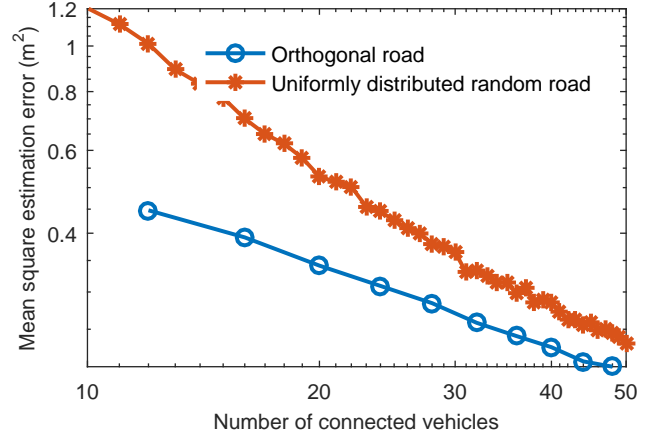}   
  \caption{Numerical results of the error decrement in the two cases of road configuration with large non-common error}
  \label{rf5}
\end{figure}
\indent Fig. \ref{rf5} shows the error decrement in the two different cases of road configuration when the non-common error variance is large. As can be observed from the figure, the slope of the decrement curve in the uniformly distributed road case is steeper than that corresponding to the orthogonal road case. This result is expected from intuition as in the former case, the road angles are more diverse, thereby, providing more constraints to correct the GNSS bias. This can also be understood from another point of view from the mathematical expression (\ref{eq8}). The estimation error is equal to the deviation of the geometric center enclosed by the road constraints. As the directions of the road angles become diverse, there is a large probability that the error in different directions cancels out. As a result, the expectation of the error become small. In contrast, if there exists some dominant directions, the expectation of the error would be large as the probability that the error cancels out becomes small.\\
\section{CONCLUSIONS}
In this paper, the impact of road configuration on the CMM localization accuracy is studied theoretically. The correlation between the mean square error of the common error estimation and the number of connected vehicles is proved analytically and shown through numerical simulation. The main results and findings are summarized:
\begin{enumerate}
\item A closed form expression of the mean square estimation error in terms of the road configuration and the non-common error is derived for the evaluation of the impact of road configuration on CMM.
 \item The mean square error of the common error estimation is inversely proportional to the logarithm of the number of connected vehicles asymptotically if the random road angles are either parallel or orthogonal.
 \item The mean square error of the common error estimation is inversely proportional to the number of connected vehicles asymptotically if the road angles obey a uniform distribution.
 \end{enumerate}




\section*{Acknowledgment}
This work is funded by the Mobility Transformation Center at the University of Michigan.
The first two authors, M. Shen and D. Zhao have equally contributed to this research.

\section*{APPENDIX}
\indent The detail with respect to the derivation of (25) and (26) is presented here.\\
\indent The direction angles of the roads are denoted by $[\theta_1,\theta_2,...,\theta_N]$. Without loss of generality, it is assumed that $0\leq \theta_1 \leq \theta_2 \leq...\leq \theta_N \leq 2\pi$. The increments of the angles are defined as
\begin{equation}
\tilde{\theta_i}=\begin{cases}
\theta_{i+1}-\theta_i & \text{for }i=1,2,...N-1\\
\theta_1-\theta_N+2\pi & \text{for }i=N\\
\end{cases}
\end{equation}
$e^2_0$ can be expressed in terms of the geometric quantities:
\begin{equation}
\begin{aligned}
e^2_0&=\frac{(\sum_{i=1}^N \frac{2}{3}w^3tan(\frac{\tilde{\theta_i}}{2})cos(\theta_i))^2}{S^2_0}\\
&+\frac{(\sum_{i=1}^N \frac{2}{3}w^3tan(\frac{\tilde{\theta_i}}{2})sin(\theta_i))^2}{S^2_0}\\
&=\frac{4w^2}{9}\frac{\sum_{i=1}^N \tan^2(\frac{\tilde{\theta_i}}{2})}{\pi^2}\\
&+\frac{4w^2}{9}\frac{\sum_{i=1}^N\sum_{j=1}^N tan(\frac{\tilde{\theta_i}}{2})tan(\frac{\tilde{\theta_j}}{2})cos(\theta_j-\theta_i)}{\pi^2}\\
&+Higher\mbox{ } Order\mbox{ } Terms
\end{aligned}
\end{equation}
\indent The probability distribution of $\tilde{\theta_i}$ can be derived:
\begin{equation}
p(\tilde{\theta}_i)=\frac{N}{\pi}(1-\frac{\tilde{\theta_i}}{\pi})^{N-1}, \mbox{ }0\leq \tilde{\theta_i}\leq \pi.
\end{equation}
\indent Accurate to the leading order:
\begin{equation}
E_{\theta}[e^2_0]=\frac{4w^2}{9\pi^2}E_{\theta}[\sum_{i=1}^N tan^2(\frac{\tilde{\theta_i}}{2})]=\frac{4w^2N}{9\pi^2}E_{\theta}[tan^2(\frac{\tilde{\theta_i}}{2})]
\end{equation}
\indent The expectation in (30) can be calculated: 
\begin{equation}
\begin{aligned}
E_{\theta}[tan^2(\frac{\tilde{\theta_i}}{2})]&=\int^{\pi}_0tan^2(\frac{\tilde{\theta_i}}{2})\frac{N}\pi(1-\frac{\tilde{\theta_i}}{\pi})^{N-1}d\tilde{\theta_i}\\
&=\int^{\frac{\pi}{\sqrt{N}}}_0(\frac{\tilde{\theta_i}}{2})^2\frac{N}\pi(1-\frac{\tilde{\theta_i}}{\pi})^{N-1}d\tilde{\theta_i}\\
&+\int^{\frac{\pi}{\sqrt{N}}}_0[tan^2(\frac{\tilde{\theta_i}}{2})-(\frac{\tilde{\theta_i}}{2})^2]\frac{N}\pi(1-\frac{\tilde{\theta_i}}{\pi})^{N-1}d\tilde{\theta_i}\\
&+\int^{\frac{\pi}{2}}_{\frac{\pi}{\sqrt{N}}}tan^2(\frac{\tilde{\theta_i}}{2})\frac{N}\pi(1-\frac{\tilde{\theta_i}}{\pi})^{N-1}d\tilde{\theta_i}\\
&+\int^{\pi}_{\frac{\pi}{2}}tan^2(\frac{\tilde{\theta_i}}{2})\frac{N}\pi(1-\frac{\tilde{\theta_i}}{\pi})^{N-1}d\tilde{\theta_i}.
\end{aligned}
\end{equation}

\indent The first term after the last equality in (31) can be integrated analytically and shown that
\begin{equation}
\int^{\frac{\pi}{\sqrt{N}}}_0(\frac{\tilde{\theta_i}}{2})^2\frac{N}\pi(1-\frac{\tilde{\theta_i}}{\pi})^{N-1}d\tilde{\theta_i}\sim \frac{\pi^2}{2N^2}.
\end{equation}\\
%


\indent The remaining three term will be shown as $o(\frac{1}{N^2})$ terms:

\begin{equation}
\begin{aligned}
&|\int^{\frac{\pi}{\sqrt{N}}}_0[tan^2(\frac{\tilde{\theta_i}}{2})-(\frac{\tilde{\theta_i}}{2})^2]\frac{N}\pi(1-\frac{\tilde{\theta_i}}{\pi})^{N-1}d\tilde{\theta_i}|\\
&<\int^{\frac{\pi}{\sqrt{N}}}_0(\frac{\tilde{\theta_i}}{2})^4\frac{N}\pi(1-\frac{\tilde{\theta_i}}{\pi})^{N-1}d\tilde{\theta_i}\\
&<\frac{\pi^2}{4N}\int^{\frac{\pi}{\sqrt{N}}}_0(\frac{\tilde{\theta_i}}{2})^2\frac{N}\pi(1-\frac{\tilde{\theta_i}}{\pi})^{N-1}d\tilde{\theta_i}\\
&=O(\frac{1}{N^3}).
\end{aligned}
\end{equation}
\begin{equation}
\begin{aligned}
&|\int^{\frac{\pi}{2}}_{\frac{\pi}{\sqrt{N}}}tan^2(\frac{\tilde{\theta_i}}{2})\frac{N}\pi(1-\frac{\tilde{\theta_i}}{\pi})^{N-1}d\tilde{\theta_i}|\\
&<\int^{\frac{\pi}{2}}_{\frac{\pi}{\sqrt{N}}}\frac{N}\pi(1-\frac{\tilde{\theta_i}}{\pi})^{N-1}d\tilde{\theta_i}\\
&=(1-\frac{1}{\sqrt{N}})^N-(\frac{1}{2})^N.
\end{aligned}
\end{equation}
\indent As
\begin{equation}
\begin{aligned}
\lim_{N\to\infty} N^2(1-\frac{1}{\sqrt{N}})^N&=\lim_{N\to\infty} N^4((1-\frac{1}{N})^N)^N\\
&=\lim_{N\to\infty} N^4e^{-N}=0,
\end{aligned}
\end{equation}
it follows that
\begin{equation}
\int^{\frac{\pi}{2}}_{\frac{\pi}{\sqrt{N}}}tan^2(\frac{\tilde{\theta_i}}{2})\frac{N}\pi(1-\frac{\tilde{\theta_i}}{\pi})^{N-1}d\tilde{\theta_i}=o(\frac{1}{N^2}).\\
\end{equation}
\begin{equation}
\begin{aligned}
&|\int^{\pi}_{\frac{\pi}{2}}tan^2(\frac{\tilde{\theta_i}}{2})\frac{N}\pi(1-\frac{\tilde{\theta_i}}{\pi})^{N-1}d\tilde{\theta_i}|\\
&<\int^{\pi}_{\frac{\pi}{2}}\frac{N}\pi(1-\frac{\tilde{\theta_i}}{\pi})^{N-3}d\tilde{\theta_i}\\
&<\int^{\pi}_{\frac{\pi}{2}}\frac{N}\pi(\frac{1}{2})^{N-3}d\tilde{\theta_i}=\frac{N}{2}(\frac{1}{2})^{N-3}=o(\frac{1}{N^2}).\\
\end{aligned}
\end{equation}
\begin{equation}
\begin{aligned}
&E_{\theta}[\frac{1}{S^2_0}tr(L^TC^TCL)]=E_{\tilde{\theta}}[\frac{1}{S^2_0}tr(L^TC^TCL)]\\
&=\sum_{i=1}^N\frac{2\sigma^2_i(E_{\tilde{\theta}}[tan^2(\frac{\tilde{\theta}_i}{2})]+E_{\tilde{\theta}}^2[tan(\frac{\tilde{\theta}_i}{2}])}{\pi^2}+H.O.T.\\
&=\frac{3\sum_{i=1}^N \sigma^2_i}{2N^2}+o(\frac{1}{N}).\\
\end{aligned}
\end{equation}



\end{document}